\documentclass[11pt,a4paper]{article}
\usepackage{jheppub}
\def\Bbb{\mathbb}
\def\BZ{\Bbb Z} 
  
\def\BN{\Bbb N}

\def\Tr{\textrm{Tr}}

\begin{document}
\bibliographystyle{jhep}
\begin{titlepage}
\renewcommand{\thefootnote}{\fnsymbol{footnote}}
\noindent
{\tt IITM/PH/TH/2009/4}\hfill
{\tt arXiv:1106.nnnn}\\[4pt]
\mbox{}\hfill 
\hfill{\fbox{\textbf{v1.0; June 7, 2011}}}
\begin{center}
\textbf{\large BKM superalgebras
from counting dyons in $\mathcal{N} = 4$\\[5pt] 
supersymmetric type II compactifications}
  \end{center}
 \bigskip \bigskip
\begin{center} 
{{\sc Suresh Govindarajan}\footnote{\texttt{suresh@physics.iitm.ac.in}}\\[2mm]
Department of Physics\\ Indian Institute of Technology
  Madras\\  Chennai 600036, INDIA\\[2mm]
  {\sc Dileep P. Jatkar}\footnote{\texttt{dileep@hri.res.in}}\\[2mm] Harish-Chandra
  Research Institute\\ Chhatnag Road,  Jhusi, Allahabad 
211019, INDIA}\\[2mm]
{\sc K. Gopala Krishna}\footnote{\texttt{krishna@mpim-bonn.mpg.de}} \\[2mm]
Max-Planck-Institut f\"ur Mathematik, \\
Vivatsgasse 7 \\
53111 Bonn, Germany
\end{center}
\bigskip

  \noindent \textbf{Abstract:} We study the degeneracy of quarter BPS dyons in $\mathcal N =4$ type
  II compactifications of string theory. We find that the genus-two
  Siegel modular forms generating the degeneracies of the quarter BPS
  dyons in the type II theories can be expressed in terms of the
  genus-two Siegel modular forms generating the degeneracies of
  quarter BPS dyons in the CHL theories and the heterotic string. This
  helps us in understanding the algebra structure underlying the
  degeneracy of the quarter BPS states. The Conway group, $Co_1$,
  plays a role similar to Mathieu group, $M_{24}$, in the CHL
  models with eta quotients appearing in the place of eta products.  We construct BKM Lie superalgebra structures for the  $\BZ_N$ (for $N=2,3,4$) orbifolds of the type II string compactified on a six-torus.

\end{titlepage}
\setcounter{footnote}{0}\baselineskip=18pt

\tableofcontents

\section{Introduction}
\label{sec:introduction}

There has been a lot of progress in understanding quarter BPS black
hole entropy in four-dimensional ${\mathcal{N}=4}$ supersymmetric
string theories\cite{Sen:2007qy, Sen:2005wa, Cardoso:2004xf}. This
understanding has been greatly aided by the possibility of having an
exact counting of the quarter BPS dyons in these string theories
\cite{Dijkgraaf:1996it, Jatkar:2005bh}. In recent times this has been
applied to a large class of four-dimensional ${\mathcal{N}=4}$
supersymmetric theories to obtain similar degeneracy formulae for the
quarter BPS dyons in these theories. In all these cases it can be
written in terms of a three-dimensional contour integral of a
genus-two Siegel modular form \cite{Sen:2007qy, Dijkgraaf:1996it,
  Jatkar:2005bh, Shih:2005uc, Gaiotto:2005hc, David:2006ji,
  Dabholkar:2006xa, David:2006yn, David:2006ru, David:2006ud,
  Dabholkar:2006bj, Banerjee:2007ub}.  The weight of these modular
forms depend on the model at hand.  In particular, the weight depends
on the number of vector multiplets in the four-dimensional
$\mathcal{N}=4$ theory. This dyon degeneracy formula, in fact, counts
an index and therefore is weakly dependent on compactification moduli.

The weak moduli dependence of the degeneracy formula manifests itself
in the form of walls of marginal stability across which certain dyons
cease to be stable and hence do not contribute to the
index\cite{Sen:2007vb, Dabholkar:2007vk,Cheng:2007ch,
  Banerjee:2008yu,Sen:2007pg}. The structure of walls of marginal
stability was understood in the axion-dilaton plane for a variety of
these models, which include toroidally compactified heterotic string
theory, $\BZ_N$ CHL models and  $\BZ_N$-orbifolds of toroidally compactified 
type II string theory.

Recently, it was shown that in the toroidally compactified heterotic
string theory as well as in $\BZ_2$, $\BZ_3$, $\BZ_4$ and $\BZ_5$ CHL
models, the dyon degeneracy formula can be written as the square of
the denominator formula for some generalized Borcherds-Ka\v c-Moody
algebra.  For each of the above models two families of algebras,
denoted $\mathcal G_N$ and $\widetilde{\mathcal G}_N$, were found. The
structure of walls of marginal stability was identified with the walls
of Weyl chambers of the corresponding Weyl groups of these
Borcherds-Ka\v c-Moody algebras\cite{Cheng:2008kt,
  Govindarajan:2008vi,Cheng:2008fc,Govindarajan:2009qt,Krishna:2010gc}.

In this paper we would like to focus our attention on the four
dimensional models with $\mathcal{N}=4$ supersymmetry that are
obtained as asymmetric orbifolds of type II string theory on $T^6$. We
shall refer to these models as `type II orbifolds'. This reflects the
fact that the chain of dualities that take one from the type IIB
string to the heterotic string in the CHL model takes one to the type
IIA string in these examples.  The type II $\BZ_N$-orbifolds (for
$N=2,3,4$) have several features in common with the CHL models.  In
particular, the structure of walls of marginal stability is identical
to the corresponding $\BZ_N$ CHL models. However, the weights of the modular
form are different.  It is therefore natural to ask which generalized
Borcherds-Ka\v c-Moody algebra encodes the dyon degeneracy formula of
type II models.

The paper is organized as follows. In section two, we provide the
background for the type II orbifolds of interest as well as the
relevant details of dyons in these models. In section three, we
explore the modular forms that generate the degeneracies of $\tfrac12$
and $\tfrac14$ BPS states. We extend our considerations to include the
$\mathbb{Z}_4$-orbifold and show that in all cases, we are able to
express the modular forms in terms of modular forms that have already
appeared in the CHL/heterotic string.  The generating function of
$\tfrac12$-BPS states are $\eta$-quotients associated with the Conway
group $Co_1$. In section 4, we explore the possibility of an algebraic
structure with these Siegel modular forms.  While we have a clear
understanding of the structure for twisted dyons (i.e. dyons invariant
under some symmetry) in the type II string, we have only a primitive
and incomplete understanding for the one counting dyons in type II
orbifolds. We conclude in section 5 with some remarks.

\section{Dyons in type II orbifolds}
\label{sec:mathc-n=4-supersymm}

\subsection{The model}

Type II string theory compactified on a six-torus has $\mathcal{N}=8$
supersymmetry in four dimensions. We will consider fixed-point free
$\mathbb{Z}_N$-orbifolds ($N=2,3,4$)  of the six-torus that
preserve $\mathcal{N}=4$ supersymmetry.  The orbifold procedure
involves splitting $T^6=T^4\times S^1\times \widetilde{S}^1$ and
choosing the action of $\BZ_N$ such that it has fixed points on $T^4$,
but this action is accompanied by a simultaneous $1/N$ shift along the
circle $S^1$. The total action of the orbifold is free, {\em i.e.}, it
has no fixed points. It thus suffices to specify the action on $T^4$.
  
As we will be moving between several descriptions of the orbifold
related by duality, we will need to specify the duality
frame. \textit{Description one} corresponds to type IIA string theory
on a six-torus with the following $\mathbb{Z}_N$ action.
Let $\omega=\exp(2\pi i/N)$ and $(z_1,z_2)$ be complex
  coordinates on $T^4$. The $\mathbb{Z}_N$ action is generated by $
  (z_1,z_2)\rightarrow (\omega z_1, \omega^{-1} z_2)$.
Our considerations generalize the $N=2,3$ orbifolds considered
in\cite{David:2006ru}.

\subsection{$\mathbb{Z}_N$ action from the NS5-brane}
 
The type II orbifolds of interest were studied originally by Sen and
Vafa who constructed dual pairs of type II orbifolds related by
U-duality\cite{Sen:1995ff}. In six-dimensional string-string duality,
the dual string is a soliton obtained by wrapping the NS5-brane on
$T^4$.  We use this observation to obtain the Sen and Vafa result by
translating the $\mathbb{Z}_N $ action on the fields in the
worldvolume of an NS5-brane wrapping $ T^4 $. Recall the fields consist
of five scalars, a second-rank antisymmetric tensor (with self-dual
field strength) in the bosonic sector and four chiral fermions. These
are the components of a single $ (2,0) $ tensor multiplet in $ 5+1 $
dimensions. We will dimensionally reduce the fields on $ T^4 $ to
obtain the fields on an effective $ 1+1 $-dimensional theory. Using
string-string duality, this theory will be that of a type IIA
Green-Schwarz string in the light-cone
gauge\cite{Dijkgraaf:1996cv,Dijkgraaf:1996hk}.
 
Let us organise the fields in terms of $ SO(4)\times SO(4)_R $ where
the first $ SO(4) =SU(2)_L\times SU(2)_R$ is from the $ T^4 $ and the
R-symmetry can be taken to be rotations about the four transverse
directions to the NS5-brane.
\begin{enumerate}
\item Four scalars, $ x^m $, are in the representation $
  (1,4_v)$. These become four non-chiral scalars upon dimensional
  reduction on the four-torus.
\item The fifth scalar and the two-form antisymmetric gauge field can
  be combined and written as $ Y_{\alpha\beta} $ and $
  Y_{\dot{\alpha}\dot{\beta}} $ where $ \alpha,\, \beta $ are $
  SU(2)_L $ spin-half indices and $ \dot\alpha,\, \dot{\beta} $ are $
  SU(2)_R $ spin-half indices.  On dimensional reduction on the
  four-torus, the $ Y_{\alpha\beta} $ become the four left-moving
  chiral bosons and the $ Y_{\dot{\alpha}\dot{\beta}} $ become four
  right-moving chiral bosons. When combined with the four non-chiral
  bosons, they become the Green-Schwarz bosons in the light-cone gauge
  of the type IIA string.
\item The fermions are $ \psi_{A\beta} $ and $ \psi_{A\dot{\beta}} $
  where $ A $ is a spinor index of $SO(4)_R $. These become the left-
  and right-moving fermions in the effective $ 1+1 $-dimensional
  theory --- these are the Green-Schwarz fermions in the light-cone
  gauge of the type IIA string.
\end{enumerate}
In the above set up, it is easy to work out transformations of fields
under $\mathbb{Z}_N$ subgroup of $SU(2)_L$.  The group element
belonging to the $\mathbb{Z}_N$ subgroup of $ SU(2)_L $ is given by
\begin{equation} 
{g_\alpha}^\beta\equiv \begin{pmatrix} \omega & 0 \\
    0 & \omega^{-1} \end{pmatrix}\ ,
\end{equation}
where $ \omega=\exp(2\pi i/N) $ for $ N=2,3,4 $. 
 
One can see that the only fields that transform under this action are
those that carry the index $ \alpha $. Thus, we see that the chiral
fermions all transform as
\begin{equation}
\psi_{A\alpha} \rightarrow {g_\alpha}^\beta\ \psi_{A\beta}\ .
\end{equation}
Thus, we see that four of the fermions pick up the phase $ \omega $ and
the other four pick up the phase $ \omega^{-1} $.  The field $
Y_{\alpha\beta} $ transforms as
\begin{equation}
Y_{\alpha\beta} \rightarrow {g_\alpha}^\gamma\ {g_\alpha}^\delta \  Y_{\gamma\delta}\ .
\end{equation}
Thus, two fields are invariant under the $ \mathbb{Z}_N $ and the
other two transform with phases $ \omega^2 $ and $ \omega^{-2}.$ All
other fields are invariant under the $ \mathbb{Z}_N $ action.
 
In the dimensional reduction of the the $ (2,0) $ theory on $ T^4 $,
the $ SU(2)_L $ fields get mapped to (say) left-movers and the $
SU(2)_R $ fields get mapped to  right-movers. Thus, we see that
the orbifold has a chiral action. In particular, the four bosons that
arise from $ Y_{\alpha\beta} $ give rise to four \emph{left-moving
  chiral bosons} and the $ \psi_{A\alpha} $ give rise to four \emph{
  left-moving chiral fermions.}

\subsection{$\mathbb{Z}_N$ action from the Poincar\'e 
polynomial}

Consider the Poincar\'e polynomial for $T^4$. Recall that two of the
one-forms pick up a phase $\omega$ while the other two pick up a phase
$\omega^{-1}$ under the $\BZ_N$ action. We incorporate this into the
Poincar\'e polynomial and obtain the action on all harmonic forms on
$T^4$.
\begin{equation}
(1-\omega x)^2 (1-\omega^{-1}x)^2=x^4-2 x^3 \omega -\frac{2
  x^3}{\omega }+x^2 \omega ^2+\frac{x^2}{\omega 
   ^2}+4 x^2-2 x \omega -\frac{2 x}{\omega }+1\ .
\end{equation}
In the above expansion, we identify even powers of $x$ with bosons in
the $1+1$ dimensional theory and odd powers with fermions. The
coefficient of a term in the polynomial gives the orbifold action on
the corresponding field. Thus six of the bosons are always periodic
and the other two have fractional moding determined by the phase.

We now present the details of the orbifold action for the
Green-Schwarz superstring that we just derived.
\begin{enumerate}
\item[{\bf [N=2]}] $\omega=-1$ implies that $\omega^2=1$. Thus, one
  has eight periodic bosons and eight anti-periodic fermions.
\item[{\bf [N=3]}] $\omega=\exp(2\pi i/3)$.  One has six periodic bosons
  and two bosons with periodicity three. Four fermions go to $\omega$
  times themselves and the other four go to $\omega^{-1}$ times
  themselves.
\item[{\bf [N=4]}] $\omega=i $. One has six periodic bosons and two
  anti-periodic bosons. Four fermions go to $\omega$ times themselves
  and the other four go to $\omega^{-1}$ times themselves.

\end{enumerate}

Thus, the second description gives rise to an asymmetric orbifold of
the type IIA string on $T^6$ and thus is analogous to CHL
compactifications of the heterotic string. In the CHL examples, recall
that the heterotic string arises from the type IIA NS5-brane wrapping
$K3$ in the place of $T^4$ that we considered.

\subsection{Tracking dyons through dualities}

Recall, the quarter BPS dyons possess charges which are
mutually non-local and therefore they do not appear in the
perturbative spectrum of the theory.   The electric charge vector
$\mathbf{Q}_e$ and the magnetic charge vector $\mathbf{Q}_m$ of a state are defined in
the second description.  One of the reasons for this choice is the
similarity of this description to the CHL description.  To see this
more explicitly, we will describe a dyonic state
in terms of a system containing $Q_5$ D5 branes wrapping $T^4\times
S^1$, $Q_1$ D1 branes wrapping $S^1$, $J$ units of momentum along
$\tilde S^1$, $-n$ units of momentum along $S^1$ and a Kaluza-Klein
monopole charged with respect to the gauge field along $\tilde S^1$.

This description is related to the second description by a chain of
duality transformations.  This chain involves first an S-duality
tranformation, which takes D-branes to NS-branes leaving all other
quantum numbers unaffected.  This is followed by a T-duality along the
circle $\tilde S^1$.  This transformation takes us from type IIB
theory to type IIA theory and replaces the Kaluza-Klein monopole by a single
NS5 brane wrapped on $T^4\times S^1$, $Q_5$ NS5 branes by $Q_5$
Kaluza-Klein monopoles along $\hat S^1$ and $J$ units of momentum
along $\tilde S^1$ is replaced by $J$ fundamental strings wrapping
$\hat S^1$, where $\hat S^1$ is a circle T-dual to $\tilde S^1$.
The $\BZ_N$-orbifold action involves $\BZ_N$-orbifold of $T^4$ and
simultaneous $1/N$ unit of shift along $S^1$.  Since the orbifolded
circle is not participating in the T-duality transformation, the
orbifold action commutes with the T-duality transformation.
Finally, one carries out string-string duality to arrive at the second
description.  The action of this string-string duality transformation
is identical to the string-string duality transformation which relates
type IIA theory compactified on K3 to heterotic string theory
compactified on $T^4$, namely, all fundamental strings are replaced by
NS5 branes and vice versa.  Thus, in the end we have $Q_1$ Kaluza-Klein
monopoles along $\hat S^1$, $-n$ units of momentum along $S^1$, $J$
NS5 branes wrapping $\hat T^4\times \hat S^1$, $Q_1$ NS5 branes
wrapping $\hat T^4\times S^1$, and a single fundamental string
wrapping $S^1$.

The second description exclusively contains description in terms of
fundamental strings, NS5 branes, Kaluza-Klein monopoles and momenta.
If we denote momenta along $S^1\times \hat S^1$ by $\mathbf{n}$,
fundamental string winding charges along them by $\mathbf{w}$ and NS5
brane, and Kaluza-Klein monopole charges by $\mathbf{N}$ and $\mathbf{W}$
respectively then the T-duality invariants constructed from these
electric and magnetic charges are
\begin{equation}
  \label{eq:19}
  \mathbf{Q}_e^2 = 2\mathbf{n}\cdot \mathbf{w}\, , \quad
  \mathbf{Q}_m^2 = 2\mathbf{N}\cdot \mathbf{W}\, , 
  \quad  \mathbf{Q}_e\cdot \mathbf{Q}_m = \mathbf{n}\cdot\mathbf{N} +
  \mathbf{w}\cdot\mathbf{W}\, . 
\end{equation}
It is easy to check that these T-duality invariants take the following
values before the orbifold action,
\begin{equation}
  \label{eq:20}
  \frac{1}{2}\mathbf{Q}_e^2 = n\, , \quad  \frac{1}{2}\mathbf{Q}_m^2 =
  Q_1Q_5\, ,\quad   \mathbf{Q}_e\cdot  \mathbf{Q}_m = J\, . 
\end{equation}
The $\BZ_N$-orbifold action commutes with the entire duality chain and
is therefore well defined in any duality frame.  It is convenient
for us to discuss it in the second description so that we can easily
read out its effect on dyonic charges.  The $\BZ_N$-orbifold acts by $1/N$
shift along $S^1$, which results in reducing the radius of the circle by a
factor of $N$.  Thus, the fundamental unit of momentum along $S^1$ is $N$
and hence momentum along $S^1$ in the orbifolded theory becomes
$n/N$.  To maintain $J$ NS5 branes transverse to $S^1$ after the
orbifold we need to start with $N$ copies of $J$ NS5 branes
symmetrically arranged on $S^1$ before orbifold.  The resulting
configuration has
\begin{equation}
  \label{eq:21}
  \boxed{
  \frac{1}{2} \mathbf{Q}_e^2 = n/N\, , \quad \frac{1}{2}
  \mathbf{Q}_m^2 = Q_1Q_5\, ,\quad 
   \mathbf{Q}_e\cdot  \mathbf{Q}_m = J\, ,
   }
\end{equation}
in the orbifolded theory.

The S-duality symmetry of this theory in the second description is
related to the T-duality symmetry in the original type IIB
description.  The $1/N$ shift along $S^1$ breaks the S-duality
symmetry of the second description to $\Gamma_1(N)$.

\section{Dyon degeneracy from modular forms}
\label{sec:dyon-degeneracy}

As mentioned in the previous section, computing the dyon spectrum is
non-trivial because dyons  do not appear in the perturbative spectrum of
string theory.  In fact, dyon counting necessarily requires computing
the degrees of freedom coming from the solitonic sector of the theory.
The dyon degeneracy formula can be obtained in two different ways,
giving rise to either the additive formula or the multiplicative one.

As shown in \cite{Jatkar:2005bh}, there are two modular forms that one
constructs -- one is the generating function of the dyon degeneracies
denoted by $\widetilde{\Phi}_k(\mathbf{Z})$ and another, denoted by
$\Phi_k(\mathbf{Z})$, is the generating function of twisted dyons in
the CHL string.\footnote{One has $\mathbf{Z} =
  \left(\begin{smallmatrix} \tau & z \\ z &
      \sigma \end{smallmatrix}\right)\in\mathcal{H}_2$ where
  $\mathcal{H}_2$ is the Siegel upper-half space. } Let us call the
corresponding modular forms in the type II orbifolds to be
$\widetilde{\Psi}_k(\mathbf{Z})$ and $\Psi_k(\mathbf{Z})$. The weight
$k$ of the Siegel modular form is given by
$$
k+2 = \frac{12}{N+1}\ ,
$$ 
when $(N+1)|12$ i.e., $N=2,3$.  For $N=4$, one has $k=1$.

\subsection{Counting electrically charged 
$\tfrac12$-BPS states}

As mentioned earlier, we will define our charges in the second
description. In this case, electrically charged states appear as
excitations of the type IIA string. In particular, the degeneracy is
dominated by the contribution from the twisted sector states. We will
compute the electrically charged states in a twisted
sector. $\tfrac12$-BPS states arise when the right-movers are in the
ground state and we allow all excitations that are consistent with
level matching as was done for the heterotic string in ref. \cite{Sen:2005ch}.

\subsubsection*{$N=1$}

As a warm-up, consider the left-movers of the type IIA string on
$T^6$. In the Ramond sector and in the light-cone gauge, one has eight
periodic bosons and periodic fermions. All oscillators, bosonic and
fermionic, have integer moding and the Witten index is given by the
product of the bosonic and fermionic contributions:
\begin{equation}
\mathcal{W}_B \times \mathcal{W}_F=
\left(\frac1{\prod_n(1-q^n)}\right)^8\times
\left(\prod_n(1-q^n)\right)^8 =1\ .
\end{equation}
This is expected as there is a perfect cancellation of bosonic and
fermionic contributions in the Witten index. Of course, the oscillator
partition function is not unity and equals
\begin{equation}
\mathcal{Z}_B \times \mathcal{Z}_F=
\left(\frac1{\prod_n(1-q^n)}\right)^8\times
\left(\prod_n(1+q^n)\right)^8 = \frac{\eta(2\tau)^8}{\eta(\tau)^{16}}\
.
\end{equation}

\subsubsection*{$N=2$}

The eight periodic bosons have integer moding and each contribute a
factor of $\eta(\tau)^{-1}$ to the Witten index while the eight
anti-periodic fermions each have half-integer moding and contribute
$\eta(\tau/2)/\eta(\tau)$.  One has
\begin{align}
\mathcal{W}_B \times \mathcal{W}_F &=
\left(\frac1{\prod_n(1-q^n)}\right)^8\times
\left(\prod_n(1-q^{n+1/2})\right)^8 \nonumber \\ 
& = \frac{\eta(\tau/2)^8}{\eta(\tau)^{16}} = \frac1{g_{\tilde{\rho}}(\tau/2)}\ ,
\end{align}
where the Frame-shape $\tilde{\rho}=1^{-8} 2^{16}$.\footnote{Frame-shapes are natural generalizations of cycles shapes  that appear in the CHL examples. (see ref. \cite{Mason:1984}). The map $g_\rho(\tau)$ that maps a Frame-shape $\rho=1^{a_1}2^{a_2}\cdots$ to the $\eta$-quotient $g_\rho(\tau)=\eta(\tau)^{a_1}\eta(2\tau)^{a_2}\cdots$ is identical to the one that appeared in the CHL strings\cite{Govindarajan:2009qt}. }  We can also
compute the twisted index for BPS states in the type II string that
are invariant under the $\BZ_2=(-1)^{F_L}$ action without accompanied
by shift along any circle.  These states contribute to the twisted
helicity supertrace $B_4^{\hat g}$\cite{Sen:2010tw}, where $\hat
g=(-1)^{F_L}$ is the generator under which these states are invariant.
In type II theory on $T^6$ this corresponds to setting all R-R fields
and R-NS fields to zero.  This means we are left with NS-NS and NS-R
sector fields in the subspace of moduli space where $(-1)^{F_L}$
symmetry is manifest.  States contributing to the twisted helicity
supertrace $B_4^{\hat g}$ belong to the elementary string states with
arbitrary excitation in the left moving sector but ground state in the
right moving sector.  These states break all 16 left moving
supersymmetries and 8 right moving supersymmetries.  However, only 8
right moving supersymmetries are invariant under $(-1)^{F_L}$, as a
result these states contribute to $B_4^{\hat g}$.  With some abuse of
language, henceforth, we will refer to these states as twisted dyons,
which are in fact $1/8$-BPS states and are counted by the twisted
helicity supertrace $B_6^{\hat g}$\cite{Sen:2010tw}.  The
$\eta$-quotient for these states is given by the S-transform i.e.,
$\tau\rightarrow -1/\tau$ of the $\eta$-quotient that we got from the
Witten index.  This leads to a second $\eta$-quotient (ignoring
numerical factors)
\begin{equation}
\frac1{g_\rho(\tau)}= \frac{\eta(2\tau)^8}{\eta(\tau)^{16}}\ ,
\end{equation}
with Frame-shape $\rho=1^{16}2^{-8}$.  Unlike the CHL examples, where
the S-transform did not change the cycle shape, we obtain \textit{a
  pair of Frame-shapes} in all the examples -- one that counts an index for
$\tfrac12$-BPS states in the orbifold of the type II string while the
other counts a twisted index for $\tfrac14$-BPS states in the type II string.

\subsubsection*{$N=3$}

The six periodic bosons have integer moding and each contribute a
factor of $\eta(\tau)^{-1}$ to the Witten index. While the two other
bosons have fractional moding of $\pm1/3$. The fermions each
have fractional moding of $\pm1/3$ and contribute
$\eta(\tau/3)/\eta(\tau)$.  One has
\begin{align}
\mathcal{W}_B \times \mathcal{W}_F &=
\frac1{\prod_n(1-q^n)^6(1-q^{n+1/3})(1-q^{n-1/3})} \times
\prod_n(1-q^{n+1/3})^4(1-q^{n-1/3})^4 \nonumber \\ 
&= \frac{\eta(\tau/3)^3}{\eta(\tau)^{9}}= \frac1{g_{\tilde{\rho}}(\tau/3)}\ ,
\end{align}
where the Frame-shape $\tilde{\rho}= 1^{-3} 3^{9}$. Counting twisted
dyons leads to a second $\eta$-quotient
\begin{equation}
\frac1{g_\rho(\tau)}= \frac{\eta(3\tau)^3}{\eta(\tau)^{9}}\ ,
\end{equation}
with Frame-shape $\rho=1^{9}3^{-3}$. 

\subsubsection*{$N=4$}

The six periodic bosons have integer moding and each contribute a
factor of $\eta(\tau)^{-1}$ to the Witten index. While the two other
bosons are anti-periodic and have half-integral
moding. The fermions each have fractional moding of $\pm1/4$ and
contribute $\eta(\tau/4)/\eta(\tau)$. One has
\begin{align}
\mathcal{W}_B \times \mathcal{W}_F &=
\frac1{\prod_n(1-q^n)^6(1-q^{n+1/2})^2} \times
\prod_n(1-q^{n+1/4})^4(1-q^{n-1/4})^4 \nonumber \\ 
&= \frac{\eta(\tau/4)^4}{\eta(\tau)^{4}\eta(\tau/2)^6}=
\frac1{g_{\tilde{\rho}}(\tau/4)}\ , 
\end{align}
where the Frame-shape $\tilde{\rho}=1^{-4} 2^{6}4^4$. Counting twisted
dyons leads to a second $\eta$-quotient
\begin{equation}
\frac1{g_\rho(\tau)}= \frac{\eta(4\tau)^4}{\eta(\tau)^{4}\eta(2\tau)^6}\ ,
\end{equation}
with Frame-shape $\rho=1^{4}2^{6}4^{-4}$.

\subsection{$\eta$-quotients, Frame-shapes and the Conway group}

The counting of $\tfrac12$-BPS states in the type II orbifold is given
by $\eta$-quotients that are associated with the Frame-shapes
$\tilde{\rho}$ while twisted $\tfrac14$-BPS states in the type II
string are associated with Frame-shapes $\rho$ as given in Table
\ref{etaquotienttable}. This nicely generalizes the corresponding
result for CHL strings where the generating functions were given by
$\eta$-products corresponding to cycle shapes.

The appearance of the $\eta$-quotients and Frame-shapes may be
understood as follows. It is known that the Conway group $Co_1$ arises
as the group of automorphisms of the algebra of chiral vertex
operators in the NS sector of the superstring\cite{Duncan:2007}. Any
symmetry of finite order of the chiral superstring \textit{must} thus
be an element of $Co_1$. It is known that the conjugacy classes of
$Co_1$ are given by Frame-shapes. It thus appears that the Conway
group plays a role analogous to the one played by the Mathieu group
$M_{24}$ with $\eta$-quotients replacing
$\eta$-products\cite{Govindarajan:2009qt}.  Is there a moonshine for
the Conway group? The $\eta$-quotients for $N=2,3$ have also appeared in the work of Scheithauer who constructed the Fake monster superalgebra as well as noted the connection with the Conway group\cite{Scheithauer:2000,Scheithauer:2004,Scheithauer:2008}.

Multiplicative $\eta$-quotients have been systematically studied by Martin and he has
provided a list of 71 such quotients -- almost all appear to be
associated to conjugacy classes of $Co_1$\cite{Martin:1996}. Table \ref{etaquotienttable} is a
subset of the list except for the ones for $N=2$. The $\eta$-quotients
for $N=2$ violate the multiplicative condition of Martin -- he
requires them to be eigenforms of all Hecke operators. The ones for
$N=2$ are not eigenforms for $T_2$ as can be easily
checked\footnote{We thank Martin for useful correspondence which
  clarified this point.}. It appears that the condition imposed by
Martin might be too strong as it excludes the $N=2$ quotient that we
obtain. It might be sufficient to require that the $\eta$-quotient be a
Hecke eigenform for all Hecke operators $T_m$ with $(m,N)=1$. The
$\eta$-quotients for $N=2,3$ have already been derived in
\cite{David:2006ru} and our results agree with the expressions given
there.

\begin{table}[h]
\newcommand\T{\rule{0pt}{2.5ex}}
\centering
\begin{tabular}{|c|c|c|c|c|c|}
\hline
 $k$ &$\widetilde{\rho}$\T & $\rho$ & $\chi\left(\begin{smallmatrix}a
     & b \\ c & d \end{smallmatrix}\right)$  &$N$ & $G$ \\[2pt] \hline 
 $2$ & $ 1^{-8} 2^{16} $\T &$ 1^{16} 2^{-8} $&  &$2$ &$\mathbb{Z}_2$ \\ \hline
 $1$ &  $ 1^{-3} 3^{9} $   & $ 1^{9} 3^{-3} $
 &$\left(\frac{-3}{d}\right)$ \T& $3$ & $\mathbb{Z}_3$  \\[3pt] \hline 
  $1$ & $1^{-4} 2^{6}4^4 $ &$ 1^{4} 2^{6}4^{-4} $&
  $\left(\frac{-1}{d}\right)$\T & $4$ & $\mathbb{Z}_4$ \\[4pt] \hline 
\end{tabular}
\caption{$\eta$-quotients with $N\leq4$:  $\widetilde{\rho}$ and $\rho$ are the pair of Frame-shapes,
  $k+2$ is the weight of the $\eta$-quotient.} 
\label{etaquotienttable}
\end{table}

\subsection{The Siegel modular forms}

We will look for genus-two Siegel modular forms, $\Psi_k(\mathbf{Z})$
and $\widetilde{\Psi}_k(\mathbf{Z})$, that have the following behavior
\begin{align} \label{condition}
\lim_{z\rightarrow0} \Psi_k(\mathbf{Z}) &\sim z^2\
g_{\tilde{\rho}}(\tau)\ g_{\tilde{\rho}}(\sigma) \ ,\nonumber \\ 
\lim_{z\rightarrow0} \widetilde{\Psi}_k(\mathbf{Z}) &\sim z^2\
g_{\rho}(\tau/N)\ g_{\tilde\rho}(\sigma)\ . 
\end{align}
\textbf{Remark:} The fractional charges here appear with the Frame-shape $\rho$ rather than the Frame-shape $\widetilde{\rho}$ that we
saw in the half-BPS counting in the orbifold theory.  This is related
to the fact that the type II dyon formula can be brought into the form
identical to the CHL dyon formula by doing the substitution $\mathbf{Q}^2_m \to
2 \mathbf{Q}^2_e$ and $\mathbf{Q}^2_e \to \mathbf{Q}^2_m/2$\cite{David:2006ru}.

We rewrite the $\eta$-quotients, $g_{\tilde{\rho}}(\tau)$,  in a
suggestive manner as 
\begin{align}
  \label{eq:3}
  g_{1^{-8} 2^{16}}(\tau) &=\frac{\eta^{16}(2\tau)}{\eta^8(\tau)} =
  \frac{\eta^{16}(2\tau)\eta^{16}(\tau)}{\eta^{24}(\tau)}\ ,\nonumber \\
 g_{1^{-3} 3^{9}}(\tau) &= \frac{\eta^9(3\tau)}{\eta^3(\tau)} =
 \frac{\eta^9(3\tau)\eta^9(\tau)}{\eta^{12}(\tau)} \ , \\
g_{1^{-4} 2^{6}4^4}(\tau)&=
\frac{\eta(4\tau)^{4}\eta(2\tau)^6}{\eta(\tau)^4}
=\frac{[\eta(4\tau)^{4}\eta(2\tau)^2\eta(\tau)^4]\times
  [\eta(\tau)^4\eta(2\tau)^4]}{\eta(\tau)^{12}}\ .\nonumber 
\end{align}
In this form, it is easy to see that the $\eta$-quotients can be written
as quotients of the $\eta$-products (or their square-roots) that appear
in the $\tfrac12$-BPS counting of the heterotic string and the CHL
orbifolds.  This naturally suggests that the Siegel modular forms
$\Psi_k(\mathbf{Z})$ can be written as quotients of the Siegel
modular forms that appear in the heterotic string and its CHL
orbifolds\cite{Jatkar:2005bh,Govindarajan:2008vi}.  Concretely, we conjecture that
\begin{align}
  \label{eq:4}
  \Psi^{N=2}_2(\mathbf{Z}) &=
  \frac{\Phi_6(\mathbf{Z})^2}{\Phi_{10}(\mathbf{Z})}\ , \\ 
  \label{eq:6}
  \Psi^{N=3}_1(\mathbf{Z}) &=
  \frac{\Delta_2(\mathbf{Z})^3}{\Delta_5(\mathbf{Z})}\ , \\ 
  \label{n4}
  \Psi^{N=4}_1(\mathbf{Z}) &=
  \frac{\Phi_3(\mathbf{Z})\Delta_3(\mathbf{Z})}{\Delta_5(\mathbf{Z})}\
  , 
\end{align}
where $\Delta_{k/2}(\mathbf{Z})^2 = \Phi_k(\mathbf{Z})$, and
$\Phi_k(\mathbf{Z})$ are 
Siegel modular forms appearing in CHL models.  It is easy to see that
the first condition in Eq.\eqref{condition} is easily
satisfied. Further, the $\widetilde{\Psi}_k(\mathbf{Z})$ can be
defined by the
S-transform\cite{Jatkar:2005bh,David:2006ud,Govindarajan:2009qt,
  Govindarajan:2010fu} 
 ($\textrm{vol}_\rho \equiv  \prod_{j=1}^N (j)^{a_j}$) 
  \begin{equation}\label{Stransform1}
 \widetilde{\Psi}_k(\mathbf{Z}) := (\textrm{vol}_\rho)^{1/2} \ \tau^{-k} \
\Psi_k(\mathbf{\widetilde{Z}})\ ,
\end{equation}
with
\[
\tilde{\tau} = -1/\tau \quad,\quad \tilde{z} = z/\tau\quad, \quad 
\tilde{\sigma} = \sigma -z^2/\tau\ .
\]
We have arrived at the relation \eqref{eq:4},\eqref{eq:6}, and
\eqref{n4} by comparing the additive seeds for Siegel modular forms
for heterotic, CHL and type II models.  We need to be careful about
the relation between the corresponding Siegel modular forms because it
could involve some phases(multiplier system); particularly so, if the
relation involves taking square roots of the Siegel modular forms or
if embedding of the genus one congruent subgroup $\Gamma_0(N)$ in the
subgroup of $Sp(2, \BZ)$ gives rise to a multiplier system for the
Siegel modular form.  In the case of $\BZ_2$-orbifold, the type II
dyon partition function is a ratio of the square of the $\BZ_2$ CHL
Siegel modular form and the Siegel modular form for the heterotic
string theory.  We are in a fortunate situation here because neither
of them have any multiplier system with respect to the subgroup of
$Sp(2,\BZ)$ which includes the genus one congruent subgroup
$\Gamma_0(2)$.  It is therefore obvious that $\Psi_2(\mathbf{Z})$ does
not have a multiplier system.  In case of $\BZ_3$ also the CHL Siegel
modular form does not have a multiplier system but the type II Siegel
modular form is not directly related to the Siegel modular form of the
$\BZ_3$ CHL model or for that matter to that of the heterotic theory.
It is a ratio of cube of $\Delta_2(\mathbf{Z})$ and
$\Delta_5(\mathbf{Z})$, where $\Delta_2(\mathbf{Z})$ is square root
of the Siegel modular form for the $\BZ_3$ CHL model and
$\Delta_5(\mathbf{Z})$ is the square root of the heterotic string
theory Siegel modular form.  Therefore $\Psi_1^{N=3}(\mathbf{Z})$, in
general, can have multiplier system under the subgroup of $Sp(2,\BZ)$
which includes $\Gamma_0(3)$.  It is therefore important to check that
the Siegel modular form $\Psi_1^{N=3}$ as well as $\Psi_1^{N=4}$, for
which genus one congruent subgroup is $\Gamma_0(4)$, do not possess a
multiplier system.  In addition to this, it is important to ensure
that the Taylor expansion of the inverse powers of these Siegel
modular forms, in terms of $\rho$, $\sigma$ and $v$ have integer
coefficients.  While it is desirable to do these checks we will derive
these results using the product formula and this derivation will
automatically ensure that both these conditions are satisfied.

\subsection{Product Representation}
\label{sec:prod-repr}

The product formula is written in terms of twisted elliptic genus of
$T^4$.  For $\mathcal{N}=4$ type II orbifolds, the $\BZ_N$ acts
simultaneously on a circle of $T^2$ by a
$1/N$ shift and on $T^4$ by $2\pi/N$ rotation.  For $N=2$, it reverses
all $T^4$ coordinates whereas for $N=3$, the orbifold action is easily
seen if we choose lattice vectors which subtend $2\pi/3$ angle with
each other.

The product formula is given in terms of Jacobi forms of weight zero
and index one, $F_{N\ II}^{(r,s)}(\tau, z)$, which are defined by
\begin{equation}
  \label{eq:7}
  F_{N\ II}^{(r,s)}(\tau,z) \equiv \frac1N \Tr_{RR;\tilde g^r} \left(\tilde g^s
(-1)^{F+\bar F} e^{2\pi i \tau L_0} e^{2\pi i {\cal J} z}\right), \qquad
r,s=0,1,\cdots , N-1\, ,
\end{equation}
where $\tilde g$ is a transformation which implements the $\BZ_N$ orbifold
transformation on the coordinates of $T^4$.  While in case of the $\BZ_2$
orbifold it changes signs of all four coordinates, for $\BZ_3$ orbifold
it generates a $2\pi/3$ rotation in a two-dimensional plane and
$-2\pi/3$ rotation in the orthogonal two-dimensional plane.  $F$ and
$\bar F$ are left and right chiral fermions in the $(4,4)$
superconformal field theory with $T^4/\BZ_N$ target space.  This
superconformal field theory has $SU(2)_L\times SU(2)_R$ R-symmetry and
${\cal J}/2$ represents the generator of the $U(1)_L$ subgroup of this
R-symmetry.

The twisted elliptic genera $F_{N\ II}^{(r,s)}(\tau, z)$ are Jacobi forms and have the following Fourier-Jacobi expansion 
\begin{equation}
  \label{eq:8}
  F_{N\ II}^{(r,s)}(\tau,z) =\sum_{b=0}^1\  \sum_{\substack{j\in 2\BZ_b, n\in \BZ/N \\ 4n-j^2 \geq -b^2}} c_b^{(r,s)}(4n -j^2)
e^{2\pi i n\tau + 2\pi i jz}\, .
\end{equation}
Explicit product formulae in terms of the Fourier-Jacobi coefficients $c_b^{(r,s)}(4n -j^2)$ for $\Psi(\mathbf{Z})$ and  $\widetilde{\Psi}(\mathbf{Z})$ as well as the CHL modular forms have been given, for instance, in ref.  \cite{David:2006ud}. We do not reproduce them here as we do not need the detailed expressions for our analysis.

For  $N=2$ and $3$, the $F^{(r,s)}(\tau,z)$ can be written as\cite{David:2006ru}
\begin{align}
  \label{eq:11}
  F_{N\ II}^{(0,0)}(\tau, z) &= 0\nonumber \\
F_{N\ II}^{(0,s)}(\tau, z) &= \frac{16}{N}\, \sin^4\left(
\frac{\pi s}{N}\right)\,  \frac{\vartheta_1\left(\tau, z+\frac{s}N
\right)\vartheta_1\left(\tau, -z+\frac{s}{N}
\right)} {\vartheta_1\left(\frac{s}{N}\right)^2}\ ,  \\
& \hspace{1.5in} \hbox{for $1\le s\le N-1$}\, , \nonumber \\ 
F_{N\ II}^{(r,s)}(\tau,z) &= \frac{4\, N}{(N-1)^2}
\, \frac{\vartheta_1\left(\tau, z+\frac{s}{N}
+\frac{r}{N}\tau
\right)\vartheta_1\left(\tau, -z+\frac{s}{N}+\frac{r}{N}\tau
\right)}{\vartheta_1\left(\frac{s}{N}+\frac{r}{N}\tau\right)^2} \ ,
\nonumber \\
& \hspace{1.5in} \hbox{for $1\le r\le N-1$,  $0\le s\le N-1$} \nonumber
\ . 
\end{align}
However, it is more instructive to write them in terms of twisted
elliptic genera of the CHL models and that of the heterotic string
theory.  Let us first note that the twisted elliptic genera of the CHL
$\BZ_N$-orbifold, $F^{(r,s)}_{N\ CHL}(\tau,z)$, can be written as\cite{Jatkar:2005bh} (for
$N=2,3,5,7$)
\begin{eqnarray}
  \label{eq:12}
  F^{(0,0)}_{N\ CHL}(\tau,z) &=& \frac{8}{N} A(\tau, z)\\
F^{(0,s)}_{N\ CHL}(\tau,z) &=&  \frac{8}{N(N+1)} A(\tau, z) -
\frac{2}{N+1} B(\tau, z) E_N(\tau) \nonumber \\
&&\qquad\qquad\qquad\qquad\hbox{\rm for}\, 1\leq s \leq (N-1) \\
F^{(r,rk)}_{N\ CHL}(\tau,z) &=& \frac{8}{N(N+1)} A(\tau, z) +
\frac{2}{N(N+1)} B(\tau, z) E_N(\frac{\tau+k}{N})\nonumber \\
&&\qquad\qquad\qquad\hbox{\rm for}\, 1\leq r \leq (N-1),\,  
0\leq k \leq (N-1),
\end{eqnarray}
where,
\begin{equation*}
  \label{eq:14}
  A(\tau, z) =  \left[ \frac{\vartheta_2(\tau,z)^2
 }{\vartheta_2(\tau,0)^2} +
\frac{\vartheta_3(\tau,z)^2}{\vartheta_3(\tau,0)^2}
+ \frac{\vartheta_4(\tau,z)^2}{\vartheta_4(\tau,0)^2}\right]\, ,
\end{equation*}
\begin{equation*}
  \label{eq:15}
  B(\tau, z) = \eta(\tau)^{-6} \vartheta_1(\tau, z)^2\, ,
\end{equation*}
and the Eisenstein series for the congruent subgroup $\Gamma_1(N)$ is
given by
\begin{equation*}
  \label{eq:16}
  E_N(\tau) = \frac{12 i}{\pi(N-1)} \, \partial_\tau \left[ \ln\eta(\tau)
-\ln\eta(N\tau)\right]= 1 + \frac{24}{N-1} \, \sum_{\substack{n_1,n_2\ge 1\\
n_1 \ne 0 \,  \textrm{mod} \, N}} n_1 e^{2\pi i n_1 n_2 \tau}\, .
\end{equation*}
For the $\BZ_2$-orbifold of the type II model, $F^{(r,s)}(\tau, z)$
can be written as\footnote{This observation was already made in a
  footnote appearing in \cite{David:2006ru} though the implication
  that the type II Siegel modular form can be written in terms of the
  CHL Siegel modular forms was not made.}
\begin{equation}
  \label{eq:17}
  F^{(r,s)}_{N=2\ II}(\tau,z) =\begin{cases}
   2 F^{(r,s)}_{N=2\ CHL}(\tau,z) -
  F^{(r,s)}_{N=1\ Het.}(\tau,z)\ , & (r,s)=(0,0) \\[3pt]
    2 F^{(r,s)}_{N=2\ CHL}(\tau,z)\ ,  & (r,s)\neq (0,0)\ .
    \end{cases}
\end{equation}
Similarly, for the $\BZ_3$-orbifold of the type II model, one has
\begin{equation}
  \label{eq:18}
  F^{(r,s)}_{N=3\ II}(\tau,z) = \begin{cases}
  \frac{3}{2} F^{(r,s)}_{N=3\ CHL}(\tau,z) -
  \frac{1}{2}F^{(r,s)}_{N=1\ Het.}(\tau,z)\ , & (r,s)=(0,0) \\[3pt]
  \frac{3}{2} F^{(r,s)}_{N=3\ CHL}(\tau,z)\ ,  & (r,s)\neq (0,0)   \ .
  \end{cases}
\end{equation}
Thus, we see that the seed for the product representation also
confirms the fact that $\Psi^{N=2}_2(\mathbf{Z})$ and
$\Psi^{N=3}_1(\mathbf{Z})$ can be written in terms of the Siegel
modular forms that appear in the CHL models and the heterotic string
theory, as stated in (\ref{eq:4}) and (\ref{eq:6}). Since, the
elliptic genera for the type II $\BZ_4$-orbifold have not been worked
out, our expression for $\Psi^{N=4}_1(\mathbf{Z})$ implies that
twisted elliptic genera can be written in terms of the CHL ones as
follows:
\begin{equation}
  F^{(0,0)}_{N=4\ II}(\tau,z) =
   F^{(0,0)}_{N=4\ CHL}(\tau,z) + \tfrac12 F^{(0,0)}_{N=2\ CHL}(\tau,z)-
  \tfrac{1}{2}F^{(0,0)}_{N=1\ Het.}(\tau,z)=0\ ,
  \end{equation}
  and for $(r,s)\neq(0,0)$, one has
  \begin{equation}
    F^{(r,s)}_{N=4\ II}(\tau,z) =  \begin{cases}
  F^{(r,s)}_{N=4\ CHL}(\tau,z) + \tfrac12 F^{(r,s)}_{N=2\
    CHL}(\tau,z)\ ,  & (r,s)= (0,0)   \mod2\\ 
  F^{(r,s)}_{N=4\ CHL}(\tau,z) \ ,  & (r,s)\neq (0,0)\mod2   \ .
  \end{cases}
\end{equation}
While we have not proved the above formulae for the $N=4$ twisted
elliptic genera, it is easy to see that it passes simple checks. For
instance, $F^{(0,0)}_{II}(\tau,z) =0$ as expected. A second check is
that $\Psi^{N=4}_1(\mathbf{Z})$ is a modular form (with character) of
a level $4$ subgroup of $Sp(2,\BZ)$ and is invariant under the
S-duality group $\Gamma_1(4)$ -- this follows from the known behavior
of the CHL modular forms\cite{Govindarajan:2009qt}.

\section{BKM superalgebras in Type II Orbifolds}
\label{sec:generalized-kac}

In their original construction, the automorphic forms constructed by
Borcherds\cite{Borcherds:1998Aut} via the singular theta lift also
happened to be related to infinite dimensional Lie superalgebras. The
automorphic form appears as the denominator identity of the BKM Lie
algebra. The infinite product representation generated by the theta
lift formed the product side of the denominator identity giving the
set of roots of the algebra and the multiplicity of these roots, while
the Fourier expansion of the automorphic form formed the sum side of
the identity which gives the Weyl group and its action on the
roots. This idea was also used by Scheithauer\cite{Scheithauer:2004},
who constructed the singular theta lifts for the elements of the
Mathieu group $M_{23}$ and $Co_1$ and showed the existence of BKM Lie
algebras for the constructed automorphic forms. The same idea was also
applied to the modular forms constructed in the CHL theory, and the
existence of BKM Lie superalgebras corresponding to these genus-two
Siegel modular forms was shown\cite{Govindarajan:2009qt,Cheng:2008kt,
  Govindarajan:2008vi,Krishna:2010gc,Govindarajan:2010fu}.

Given that the Siegel modular forms generating the dyonic degeneracies
in the type II models can be expressed in terms of ratio of the Siegel
modular forms that appear in the description of the quarter BPS dyons
in the heterotic and $\BZ_N$-orbifolded CHL strings, the possibility
arises that one can construct BKM Lie superalgebras corresponding to
the type II modular forms also. We explore this possibility in this
section. Before we do so, however, we remark on an important point
from the corresponding constructions in the CHL theory as well as
Borcherds's work.

In the case of $\BZ_N$-orbifolded CHL theories with $N>1$, there
exists more than one cusp, and the modular form has more than one
product expansion corresponding to each of the cusps. Thus,
interpreted as the denominator identity of an infinite dimensional Lie
algebra, each of the different expansions corresponds to
different BKM Lie superalgebras. The modular forms at the different
cusps correspond (in some sense) to twisted and untwisted counting of
dyonic degeneracies. This phenomenon is not particular to the CHL
models. An example in a context different from the CHL theory is that
of the denominator identity of the fake monster superalgebra which has
two completely different algebras corresponding to the cusps at level
$1$ and $2$ (see\cite{Borcherds:1998Aut} Example 13.7). Again, the
algebras at the two cusps correspond to twisted and untwisted counting
of states. The level $1$ cusp gives a BKM Lie superalgebra for
superstrings on a $T^{10}$, while the level $2$ cusp corresponds to a
twisted denominator formula corresponding to an automorphism that is
$1$ on the Bosonic elements and $-1$ on the Fermionic
ones. Remarkably, the two algebras are completely different from each
other, with different Weyl vectors, Weyl groups (the level $2$ algebra
has a trivial Weyl group), and different real simple roots (the level
$1$ algebra has an infinite number of real simple roots, while the
level $2$ algebra has no real simple roots at all). Thus we see that
the BKM Lie superalgebras corresponding to the expansions about the
different cusps are quite different from each other.

Coming back to the type II models, we now try to find the algebraic
structures, if any, occurring in the type II models. The method we
will use is the general prescription outlined
in\cite[see Appendix D.1]{Govindarajan:2008vi} for understanding
the algebra structure from the expansion of the modular form. The
modular forms occurring in the type II models seem less amenable to an
algebraic interpretation than their CHL counterparts. Where the
modular forms occurring in the CHL theory seemed to be related to two
families of BKM Lie superalgebras, the modular forms in the type II
theories have no such obvious forms. Nevertheless, we have guidance from
the fact that the modular forms of the type II theory are related to
the ones occurring in the CHL theory and this can help us derive some
algebraic structure from the corresponding structures in the CHL
theory. However, the extent of this insight gets limited by the fact
that the modular forms of the type II theory are given as quotients of
the CHL modular forms, and this does not always translate into a
straightforward interpretation of the characters of representations.

For the $N=2$ orbifold, the modular forms at the two cusps, given in
terms of the CHL modular forms are
\begin{equation}
\label{n2}
\Psi_2(\mathbf{Z}) = \frac{\Phi_6(\mathbf{Z})^2}{\Phi_{10}(\mathbf{Z})}, \quad
\textrm{ and }   \quad \widetilde{\Psi}_2(\mathbf{Z}) =
\frac{\widetilde{\Phi}_6(\mathbf{Z})^2}{\Phi_{10}(\mathbf{Z})} \ . 
\end{equation}
Using the CHL examples as a guide, the modular forms relevant for the BKM Lie superalgebras are the
square roots of the above modular forms and are given as
\begin{equation}
\label{n2root}
\Xi_1(\mathbf{Z}) =
\frac{\Delta_3(\mathbf{Z})^2}{\Delta_{5}(\mathbf{Z})}, \quad 
\textrm{ and }   \quad \widetilde{\Xi}_1(\mathbf{Z}) =
\frac{\widetilde{\Delta}_3(\mathbf{Z})^2}{\Delta_{5}(\mathbf{Z})} \ . 
\end{equation}

Let us start with the modular form $\Xi_1$ to look for a BKM Lie
superalgebra, if any, associated to it. Since $\Xi_1$ is given as the
ratio of $\Delta_3$ and $\Delta_5$, both of which have the BKM Lie
superalgebras, $\mathcal G_2$ and $\mathcal G_1$ respectively,
associated to them, we look at the corresponding algebras, $\mathcal
G_2$ and $\mathcal G_1$. 

The algebras $\mathcal G_1$ and  $\mathcal G_2$ have  the same set of real
simple roots as $\mathcal G_1$ -- $\alpha_1, \alpha_2$ and
$\alpha_3$ with Cartan matrix\cite{Nikulin:1995,Govindarajan:2008vi}
\begin{equation}
A_{1,II}\equiv \begin{pmatrix} ~~2 & -2 &-2\\ -2 & ~~2&-2\\-2&-2&~~2 
\end{pmatrix}\ .
\end{equation}
The difference between the two algebras is in their
imaginary roots. For instance, the multiplicity of light-like roots, $t\eta_0$, is given by the formula ($\eta_0$ is a primitive light-like root and $t\in \BZ_{>0}$)
\begin{equation}\label{lightlikemult}
1-\sum_{t\in \BN} m(t \eta_0)\ q^n 
= \prod_{n\in\BN} (1-q^n)^{\tfrac{k-4}2} (1-q^{Nn})^{\tfrac{k+2}2},
\end{equation}
where $k$ for $\mathcal G_1$ and $\mathcal G_2$ is $10$ and $6$
respectively. Now a quotient of the modular forms suggests the real
simple roots of the resulting algebra will remain same, since two
copies of the roots exist in the numerator $\Delta_3^2$, while one
copy of the roots is cancelled by the presence of $\Delta_5$. This
seems to suggest the algebra corresponding to $\Xi_1$ will be one with
the same real simple roots as $\mathcal{G}_1$. 
The multiplicity of the light-like roots can be guessed from the
eq. \eqref{lightlikemult} and using the quotient form of $\Xi_1$ in
terms of $\Delta_3$ and $\Delta_5$. The multiplicity of the light-like
roots, $t\eta_0$, for $\mathcal G_{II,2}$ is given by
\begin{equation}\label{Z2lightlikemult}
1-\sum_{t\in \BN} m(t \eta_0)\ q^n 
= \prod_{n\in\BN} (1-q^n)^{-8} (1-q^{2n})^{8} \ .
\end{equation}
Thus, we see that one can associate a BKM Lie superalgebra, $\mathcal
G_{II,2}$, to the modular form $\Xi_1$.

Turning to the modular form
at the other cusp, we see it can be written as the quotient of the
modular forms $\widetilde{\Delta}_3$ and $\Delta_5$. We might hope to
repeat the above success in obtaining $\mathcal{G}_{II,2}$ to find a
BKM Lie superalgebra associated to the modular form
$\widetilde{\Xi}_1$, but it is not so simple at the other cusp. The algebra $\widetilde{\mathcal{G}}_2$ (associated with $\widetilde{\Delta}_3$)
has four real simple roots, of which two are different from that
occurring in $\mathcal{G}_1$. In particular, one of the real simple roots of $\mathcal{G}_1$ will appear as a pole in the denominator identity. It is potentially a fermionic simple root. Thus, considering only the real simple roots of BKM Lie superalgebra that might be associated with the modular form $\widetilde{\Xi}_1$, we expect to see four bosonic (even) real simple roots (those appearing in $\widetilde{\mathcal{G}}_2$) and a fermionic (odd) real simple root (the one contributing to the pole). Even in the finite case, the structure of Lie superalgebras is technically more complicated than the one for classical Lie algebras.  Gritsenko and Nikulin, in their analysis of BKM Lie superalgebras consider only those superalgebras \textit{without} odd real simple roots i.e., these superalgebras only contain odd imaginary simple roots\cite[see Appendix]{Nikulin:1995}. The BKM Lie superalgebras that appear in the CHL dyon counting as well as those in Scheithauer's Fake Monster Lie superalgebra belong to this category\cite{Scheithauer:2000}. We defer a detailed discussion for the future.

It is instructive to write eq.(\ref{n2}) in the following form
\begin{equation}
  \label{eq:9}
  \Phi_6(\mathbf{Z})\times \Phi_6(\mathbf{Z})=\Psi_2(\mathbf{Z})\times
  \Phi_{10}(\mathbf{Z})\, ,
\end{equation}
and similarly for $\tilde \Psi_2(\mathbf{Z})$.  Written in this form we
can interpret it as decomposition of product of two characters
belonging to trivial representations of the BKM algebra corresponding
to two copies of $N=2$ CHL in terms of a character of trivial
representation of the BKM algebra corresponding to the heterotic
theory and $N=2$ type II model.  This is quite analogous to the way
coset models were constructed using affine Ka\v c-Moody symmetry, and
in particular to the way characters of the coset models were derived.
It is worth pointing out that the following sequence of Cartan matrices
shows relation of the BKM algebra associated with $\tau=i\infty$ cusp
of the Heterotic string and its CHL analogs to simple and affine SU(2)
Lie algebra, 
\begin{equation}\label{cartanmatrices}
  A_1=\begin{pmatrix} 2 \end{pmatrix} \hookrightarrow
  A_1^{(1)}=\begin{pmatrix} ~~2 & -2 \\ -2 & ~~2 \end{pmatrix} \hookrightarrow
  A_{1,II}\equiv \begin{pmatrix} ~~2 & -2 &-2\\ -2 & ~~2&-2\\-2&-2&~~2 
\end{pmatrix}.
\end{equation}
With this close analogy it is tempting to speculate that the type II
models possess a BKM generalization of the affine Goddard-Kent-Olive type
coset symmetry\cite{GKO}.  It would be interesting to explore this
relation further.  It will also enable us to give an unified description of 
type II orbifold models as BKM cosets.

We next come to the case of the $\mathbb Z_3$-orbifold. The modular
forms generating the degeneracy of the $\tfrac14$-BPS states for the
$\mathbb Z_3$-orbifold can be written, in terms of the CHL modular
forms as
\begin{equation}
\Psi^{N=3}_1(\mathbf{Z}) =
\frac{\Delta_2(\mathbf{Z})^3}{\Delta_5(\mathbf{Z})}\ , \quad 
\widetilde{\Psi}^{N=3}_1(\mathbf{Z}) =
\frac{\widetilde{\Delta}_2(\mathbf{Z})^3}{\Delta_5(\mathbf{Z})} \ . 
\end{equation}
We cannot take a square-root as we did for the $N=2$ case as it leads
to modular forms with non-integral coefficients. Nevertheless, we
observe that the real bosonic simple roots that appear in the
denominator of $\Psi_1(\mathbf{Z})$ are cancelled by one
$\Delta_2(\mathbf{Z})$. It leaves behind two sets of three real simple
roots with Cartan matrix $A_{1,II}$ -- this is the Cartan matrix of a
rank three Lorentzian Lie algebra\cite{Nikulin:1995}. Thus, the
modular form appears to be the product of at least two BKM Lie
superalgebras both of which are inequivalent automorphic extensions of
the Lorentzian Lie algebra with Cartan matrix $A_{1,II}$\cite{Nikulin:1995}. Physically, it means that the
walls of marginal stability of twisted dyons is identical to that of
the $N=2$ orbifold. The same conclusion is obtained for the $N=4$
modular form $\Psi^{N=4}_1(\mathbf{Z})$.  However, we do not have a
complete understanding of the algebraic structure for modular forms that
generate degeneracies of $\tfrac14$-BPS in the type II orbifolds,
i.e., for the $\widetilde{\Psi}^{N=3}_1(\mathbf{Z})$ and
$\widetilde{\Psi}^{N=4}_1(\mathbf{Z})$ as in the $N=2$ example.

Thus, we see that for the type II models one can associate a BKM Lie
superalgebra structure to the expansion of the modular forms
generating the degeneracy of twisted $\tfrac14$-BPS states along the same
lines as for the algebras in the CHL models. However, only the
expansion about one of the cusps admits an algebra structure, while it
is not very clear if an algebra exists about the other cusp.

\section{Conclusion}

In this work we have attempted to understand the degeneracy of the
quarter BPS dyons in $\mathbb Z_N$-orbifolds of $\mathcal N=4$ type II
compactification of string theory by studying the genus-two Siegel
modular forms generating the degeneracies of these states. We see that
the Siegel modular forms for the $\mathbb Z_N$-orbifolds in the type II
compactifications are expressible in terms of the CHL Siegel modular
forms.  This relation exists for both the generating functions of the
half as well as the quarter BPS states of the type II theory. Using
this we construct the algebraic structures underlying these
degeneracies. The algebras underlying twisted dyons in the type II string are all inequivalent automorphic corrections to the rank three Lorentzian Kac-Moody algebra with Cartan matrix $A_{1,II}$ consistent with the physical requirement that they have identical walls of marginal stability.

\bigskip

\noindent \textbf{Acknowledgments:} We thank Y. Martin and U. Ray for useful conversations as well as correspondence. One of us, DPJ, thanks IITM for hospitality during a visit where some of the work reported was carried out.

\bibliography{master}
\end{document}